\documentclass{aastex61}
\usepackage{graphicx}
\usepackage{longtable}
\usepackage{multirow}
\usepackage{amsmath}
\usepackage{url}
\usepackage{amsmath,bm}
\usepackage{lipsum}
\usepackage{rotating}

\date{\today}

\begin{document}

\title{A New Method to Test the Einstein's Weak Equivalence Principle}

\author{Hai Yu}
\affiliation{School of Astronomy and Space Science, Nanjing University, Nanjing 210093, China}

\author{Shao-Qiang, Xi}
\affiliation{School of Astronomy and Space Science, Nanjing University, Nanjing 210093, China}

\author{Fa-Yin Wang}
\altaffiliation{fayinwang@nju.edu.cn}
\affiliation{School of Astronomy and Space Science, Nanjing University, Nanjing 210093, China}
\affiliation{Key Laboratory of Modern Astronomy and Astrophysics (Nanjing University), Ministry of Education, Nanjing 210093, China}

\begin{abstract}
The Einstein's weak equivalence principle (WEP) is one of the
foundational assumptions of general relativity and some other
gravity theories. In the theory of parametrized post-Newtonian
(PPN), the difference between the PPN parameters $\gamma$ of
different particles or the same type of particle with different
energies, $\Delta \gamma$, represents the violation of WEP. Current
constraints on $\Delta \gamma$ are derived from the observed time
delay between correlated particles of astronomical sources. However,
the observed time delay is contaminated by other effects, such as
the time delays due to different particle emission times, the
potential Lorentz invariance violation and none-zero photon rest
mass. Therefore, current constraints are only upper limits. Here we
propose a new method to test WEP based on the fact that the
gravitational time delay is direction-dependent while others are not. This is the first
method which can naturally correct other time delay effects. Using
the time delay measurements of BASTE gamma-ray burst (GRB) sample
and the gravitational potential of local super galaxy cluster
Laniakea, we find that the constraint on $\Delta \gamma$ of
different energy photons can be as low as $10^{-14}$. In the future,
if more gravitational wave events and fast radio bursts with much
more precise time delay measurements are observed, this method can
give a reliable and tight constraint on WEP.
\end{abstract}

\section{Introduction}\label{sec:intro}
As one of the basic assumptions of general relativity and many other
gravity theories, the Einstein's weak equivalence principle (WEP) is
an important physical principle needed to be tested. It states that
the trajectories of any freely falling, uncharged bodies are
independent of their energy,  composition, or internal structure. In
the theory of parametrized post-Newtonian (PPN), the PPN parameter
$\gamma$ of different particles or the same type of particle with
different energies (hereafter, ``different particles'' represents
both of cases) should be the same
\citep{Will2014LRR....17....4W,Wei2015PhRvL.115z1101W}. Therefore,
comparing the $\gamma$ parameters of two different particles can be
used to test whether the WEP is valid. The first practicable
astronomical method to test the WEP was proposed by Shapiro in 1964
by measuring the time delay, which is lately called Shapiro
time delay, between transmission of radar pulses towards the inner
planets and detection of the echoes
\citep{Shapiro1964PhRvL..13..789S}. The first application of this
method on astrophysical source was achieved in 1988
\citep{Krauss1988PhRvL..60..176K,Longo1988PhRvL..60..173L}.
They used the different arrival times of photons and
neutrinos from SN1987A to test the WEP and found the violation of
WEP, which means the difference between the $\gamma$ parameters of
photons and neutrinos, should be less than $0.2\%-0.5\%$. Recently,
more works on this topic have been done based on the time delay of
different particles from various of astrophysical sources, such as
the photons in different energy bands of gamma-ray bursts (GRBs)
\citep{Gao2015ApJ...810..121G}, radio signals at different frequency
bands of fast radio bursts (FRBs)
\citep{Wei2015PhRvL.115z1101W,Nusser2016ApJ...821L...2N,Zhang2016arXiv160104558Z}
and the Crab pulsar \citep{Yang2016PhRvD..94j1501Y}, a PeV-energy
neutrino event associated with a giant flare of the blazar PKS
B1424-418 \citep{Wang2016PhRvL.116o1101W}, polarized photons
\citep{2017MNRAS.469L..36Y} and gravitational wave (GW) sources
\citep{Wu2016PhRvD..94b4061W,Kahya2016PhLB..756..265K}.

It should be pointed out that the observed time delay of two
different particles might be contaminated by many effects, such as
the intrinsic time delay $\Delta t_{\rm int}$ due to the
non-simultaneous emitting of particles, the time delay $\Delta
t_{\rm LIV}$ caused by the potential Lorentz invariance violation
(LIV), $\Delta t_{\rm spec}$ caused by potential non-zero rest mass
of photon, and $\Delta t_{\rm DM}$ caused by the contribution of
dispersion of photons. In general, the observed time delay of two
different particles can be expressed as
 \begin{equation}\label{eq:deltaT_obs}
  \Delta t_{\rm obs}=\Delta t_{\rm int}+\Delta t_{\rm LIV}+\Delta t_{\rm spec}+\Delta t_{\rm DM}+\Delta t_{\rm gra},
 \end{equation}
where $\Delta t_{\rm gra}$ is the relative Shapiro time delay which
corresponds to the difference in arrival time of two different
particles due to the gravitational potential. In PPN theory, it can
be formulated as
 \begin{equation}\label{eq:relaShapiroDelay}
  \Delta t_{\rm gra} = \frac{\gamma_1-\gamma_2}{c^3}\int_{\mathbf{r}_e}^{\mathbf{r}_0}\Psi(\mathbf{r^\prime}){\rm d}\mathbf{r}=\Delta\gamma f(\mathbf{r}),
 \end{equation}
where $\gamma_1$ and $\gamma_2$ are the PPN parameters of the two
different particles, $r_e$ and $r_0$ are the positions of source and
observer respectively, $c$ is the speed of light, and $\Psi(r)$ is
the gravitational potential and
$f(\mathbf{r})=\frac{1}{c^3}\int_{\mathbf{r}_e}^{\mathbf{r}_0}\Psi(\mathbf{r}){\rm
d}\mathbf{r}$.

In previous literature, all of the constraints on WEP were obtained
using the observed time delay directly
\citep{Krauss1988PhRvL..60..176K,Longo1988PhRvL..60..173L,Gao2015ApJ...810..121G,Wei2015PhRvL.115z1101W,Nusser2016ApJ...821L...2N,
Zhang2016arXiv160104558Z,Yang2016PhRvD..94j1501Y,Wang2016PhRvL.116o1101W,Wu2016PhRvD..94b4061W,Kahya2016PhLB..756..265K}.
Therefore, some assumptions on the nuisance time delays caused by
other effects should be made. In general, they assumed that the time
delays $\Delta t_{\rm LIV}$, $\Delta t_{\rm spec}$ were negligible
and $\Delta t_{\rm DM}$ could be omitted for high-energy photons.
What's more, it must be assumed that the sign of $\Delta t_{\rm
int}$ and $\Delta t_{\rm gra}$ are the same. Even with these
assumptions, they can only give an upper limit of the violation of
WEP. Recently, Yu and Wang proposed a new method based on
the strongly lensed transients, which can solve the intrinsic time
delay problem \citep{Yu2018arXiv180101257Y}.

To correct the nuisance effects in observed time delay, we propose a
new robust method to test WEP based on the global fitting of time
delay measurement of cosmic transients. In general,
gravitational time delay is direction-dependent while others are
not. We will give a detailed introduction to our method in section
\ref{sec:method} and then  we use the time delay data of GRBs in
BATSE sample to constrain the potential violation of WEP in section
\ref{sec:application}. Finally, we will give some discussion on the
uncertainty and efficiency of our method and also a short summary of
our work in section \ref{sec:discussion}. Throughout the paper, we
use flat-$\Lambda$CDM with $H_0=70\,$km/s/Mpc and $\Omega_m=0.3$ as
the fiducial cosmology model.

\section{The Method}\label{sec:method}
From equation (\ref{eq:deltaT_obs}), several effects contribute to
the observed time delay measurement. Because the strong coupling of
these effects and the lack of understanding about the physical
mechanism of the source and the physical properties of the particle
path, it is difficult to extract the $\Delta t_{\rm gra}$ from
observed time delay. Interestingly, the different statistical
properties of the time delays caused by different effects can give
us a good chance to correct the contaminations in observed time
delay measurement. Therefore, we need to analyze these effects in
detail.

Considering a kind of cosmic transients, such as GRBs, FRBs or GW
events, there is a time delay measurement between the light curves
in two energy bands (hereafter, we will use GRB as an example).
Since the cosmological principle states that the distribution of
matter in the universe is isotropic and homogeneous at large scale,
the distribution of GRBs should be also independent of the
direction. Therefore, one can naturally assume that the statistical
properties, including the intrinsic time delay between two energy
bands $\Delta t_{\rm int}$, of GRBs are independent of the
direction. LIV is one of the quantum gravity effect and many quantum
gravity theories predict that the LIV will happen at high energy
band. Since high-energy photons may interact with the foamy
structure of space time at very small scale, the speed of
high-energy photons will depend on their energy
\citep{Amelino-Camelia2013LRR....16....5A}. Therefore, the time
delay $\Delta t_{\rm LIV}$ only depends on the energy bands of
photons and the distance of the source. Similarly, if photons have
non-zero rest mass, the speed of light in vacuum is no longer a
constant but depends on the energy of photon. Therefore, the time
delay $\Delta t_{\rm spec}$ is also independent of source direction.

The Shapiro time delay is related to the distribution of
gravitational potential, the value of $\Delta t_{\rm gra}$ depends
on the direction of the source if the observer is not at the center
of a spherically symmetric gravitational potential. Since our earth
is not at the center of Milky Way and is also not at the center of
the local super galaxy cluster which is called Laniakea potential
\citep{Tully2014Natur.513...71T}, the $\Delta t_{\rm gra}$ of GRBs
from different directions should be different if there is any
violation of WEP. Actually, the $\Delta t_{\rm gra}$ includes two
parts, one of which is caused by the Laniakea potential $\Delta
t_{\rm gra,L}$ and the other is caused by the potential fluctuations
of the large scale structure $\Delta t_{\rm gra,LSS}$ which can only
contribute to the uncertainty but not the mean value of $\Delta
t_{\rm gra}$ \citep{Nusser2016ApJ...821L...2N}. In addition, based
on the cosmological principle, one can also assume that the $\Delta
t_{\rm
    gra,LSS}$ is statistically isotropic while the $\Delta t_{\rm
    gra,L}$ depends on the direction of the source. Besides, the time
delay $\Delta t_{\rm DM}$ is also dependent on the direction of
source since the distribution of free electrons is not spherically
symmetric. However, this effect is significant only for the radio
signal but negligible for high-energy photons, GWs and neutrinos.
Therefore, if we use the time delay among high energy photons, GWs
or neutrinos, the term $\Delta t_{\rm DM}$ can be omitted.

According to the analysis above, the observed time delay can be
divided into two parts. The first part of observed time delay is
$\Delta t_{\rm gra,L}$, which depends on the direction of source.
While the second part is $\Delta t_{\rm other}$ which contains all
direction-independent components, $\Delta t_{\rm gra,LSS}$, $\Delta
t_{\rm int}$, $\Delta t_{\rm LIV}$, and $\Delta t_{\rm spec}$ in
observed time delay. Therefore, the expectation of the observed time delay can be
expressed as
\begin{equation}\label{eq:fitFun}
\Delta t_{\rm th}=\Delta\gamma f(\mathbf{r})+\Delta t_{\rm other},
\end{equation}
where $\mathbf{r}$ is the position of the source and $\Delta t_{\rm
    other}=\Delta t_{\rm gra,LSS}+\Delta t_{\rm int}+\Delta t_{\rm
    LIV}+\Delta t_{\rm spec}$. If there is a large sample of time delay
measurement, for each GRB$_i$ there is a time delay measurement
$\Delta t_{{\rm obs},i}=\Delta\gamma f(\mathbf{r}_i)+\Delta t_{{\rm
        other},i}$. The $\Delta t_{{\rm other},i}$ depends on the properties
such as the redshift, explosion mechanism, and all the factors on
the line of sight of the GRB$_i$. If there are lots of GRBs
in a direction bin, one can figure out the mean value $\langle\Delta
t_{\rm other}\rangle$ and standard deviation $\sigma_{\rm other}$ of
the $\Delta t_{{\rm other},i}$ of all the GRBs in that bin. Based on
the cosmological principle, one can naturally assume that all the
factors, on which the $\Delta t_{{\rm other},i}$ depends, are
statistically independent of the direction. Therefore, the $\Delta
t_{{\rm other},i}$ should be followed by a same distribution. Based
on the central limit theorem, it is natural to assume that the
$\langle\Delta t_{\rm other}\rangle$ should follow a Gaussian
distribution $\aleph(\mu,\sigma)$, where $\mu$ is the mean of the $\Delta
t_{\rm other}$ and $\sigma=\sigma_{\rm other}/\sqrt{N}$ with $N$ is
the number of data in that bin. Therefore, it has
\begin{equation}\label{eq:fitFun2}
\langle\Delta t_{\rm th}\rangle_{\rm bin}=\Delta\gamma
f(\mathbf{r})+\langle\Delta t_{\rm other}\rangle.
\end{equation}
Then we can constrain the parameters $\Delta\gamma$,
$\langle\Delta t_{\rm other}\rangle$ by fitting equation
(\ref{eq:fitFun2}) with the time delay measurement data of GRBs in
our sample. If there is really some violation of WEP, our method can
give a non-zero value of $\Delta\gamma$.

\section{Constraining WEP from GRB data}\label{sec:application}
GRBs are the most violent explosions in the universe since Big Bang
\citep{Meszaros2006RPPh...69.2259M,Kumar2015PhR...561....1K}. Because
they can be detected at high redshifts due to their high luminosity,
GRBs are regarded as a very important tool to investigate the early
universe \citep{Wang2015NewAR..67....1W}. As one of the most
important astrophysical explosion phenomenon, several satellites
were designed and launched, such as BATSE, Swift and Fermi, to
observe GRBs. Recently, the time delays of GRBs in different energy
bands are used to test the WEP \citep{Gao2015ApJ...810..121G} and
also LIV \citep{Wei2017ApJ...834L..13W}. In this work, we will use
the time delay measurement of BATSE GRB sample with our new method
to constrain the WEP.

\begin{figure}
    \centering
    \includegraphics[width=0.50\textwidth]{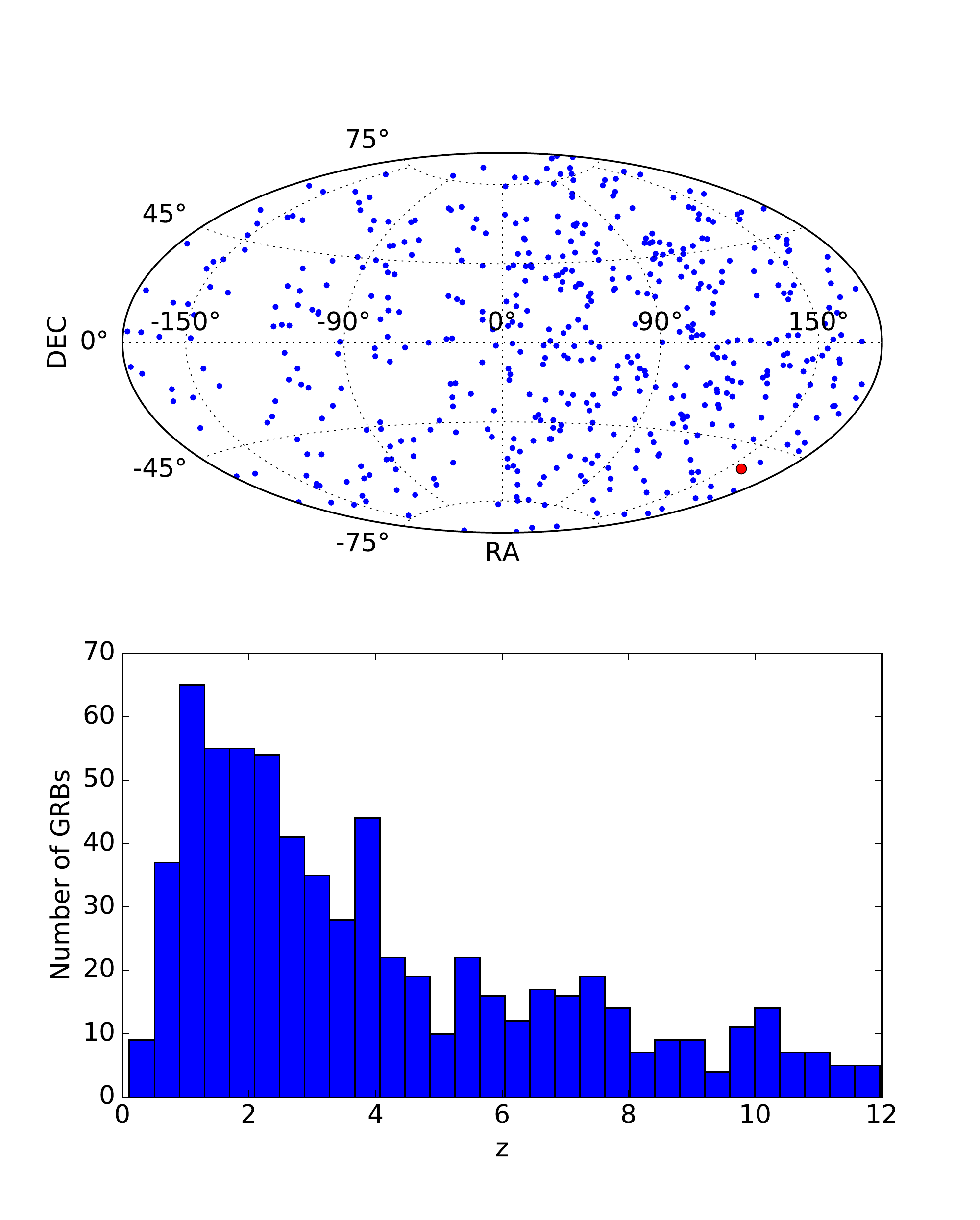}\\
    \caption{The distributions of directions and redshifts of GRBs in our sample. In the top panel,
    the blue points are the GRBs in our sample and the red point is the center of Laniakea.}\label{fig:distribution}
\end{figure}

BATSE is one of the most famous GRB detectors. It observed 2702 GRBs
in 9 years. \cite{Hakkila2007ApJS..169...62H} developed a catalog of
spectral time lags of BATSE GRBs using the BATSE 64 ms
discrimination data. Since BATSE
has 4 energy bands, Ch1: 25-60 keV, Ch2: 60-110 keV, Ch3: 110-325
keV and Ch4: $>$325 keV, there are at most 6 time delay measurement
for each GRB. \cite{Hakkila2007ApJS..169...62H} calculated 8552 time
delay measurements in total. However, since the time delay $\Delta
t_{\rm gra}$ depends on the distance of the source while there are
only few BATSE GRBs with observed redshifts, we have to use the
pseudo-redshifts of those GRBs. Fortunately, the pseudo-redshifts of
689 BATSE GRBs have been derived based on the spectral peak
energy-peak luminosity relation \citep{Yonetoku2004ApJ...609..935Y}.
We also use the BATSE 5B GRB Spectral Catalog
\citep{Goldstein2013ApJS..208...21G} to search the directions of
GRBs. According to these three catalogs, we choose 668 GRBs in our
sample. The distributions of directions and pseudo-redshifts of these GRBs
are shown in figure \ref{fig:distribution}. In the top panel of
figure \ref{fig:distribution}, the blue points represent the GRBs in
our sample and their distribution is nearly uniform. Because of the
very low signal-noise ratio, some GRBs do not have 6 time delay
measurements. The sample of the time delay measurements between
different energy channels can be found in table \ref{tab:data}.

To use the approximation that the $\langle\Delta t_{\rm
other}\rangle$ have same expectation for all directions, we need to
check whether the time delay of GRBs follows a same distribution for
different areas on the sky. We separate the whole sky into four
parts with same area and they are NE (RA$\le12^h$ and DEC$\ge$0), NW
(RA$>12^h$ and DEC$\ge$0), SE (RA$\le12^h$ and DEC$<$0) and SW
(RA$>12^h$ and DEC$<$0) respectively. Because the $\langle\Delta
t_{\rm other}\rangle$ contributes most part of the total observed
time delay $\Delta t_{\rm obs}$, we just use the two-sample
Kolmogorov-Smirnov (KS) test to check whether the
distributions of the $\Delta t_{\rm obs}$ of the four subsamples are
the same as the whole sample or not. The null hypothesis of the
two-sample KS test is that the samples are drawn from the same
distribution. We list their average and standard derivative values
and all the p-values in table \ref{tab:statistic} from which we can
see that all the p-values are too large to reject the null
hypothesis. Therefore, the approximation that the $\langle\Delta
t_{\rm other}\rangle$ have same expectation for all directions is
reliable.

To calculate the theoretical time delay $\Delta t_{\rm gra,L}$, one
has to know the gravitational potential $\Psi(\mathbf{r})$. In
general, the photons emitted from GRBs will pass through the
gravitational potentials of cosmic large scale structure, the local
super galaxy cluster Laniakea and Milky Way. However, in this work,
we will pay more attention to the dependence of $\Delta t_{\rm gra,L}$
on the positions of GRBs. Because the gravitational potential of
cosmic large scale structure can be regarded as an isotropic one, we
can simply treat this effect $\Delta t_{\rm gra,LSS}$ in the $\Delta t_{\rm other}$ term.
Because the total mass of Laniakea is about $10^{17} {\rm{M_\odot}}$
which is about $10^5$ times heavier than the Milky Way
\citep{Tully2014Natur.513...71T}, the effect of Laniakea is much more
important. Therefore, we here consider the potential anisotropic term
$\Delta t_{\rm gra,L}$ is meanly caused by the gravitational
potential of Laniakea super galaxy cluster. Adopting a Keplerian
potential for Laniakea, we have \citep{Longo1988PhRvL..60..173L}
\begin{equation}\label{eq:fr}
f(\mathbf{r})=\frac{GM_L}{c^3}\times\ln\{\frac{[d+(d^2-b^2)^{1/2}][r_L+s_n(r_L^2-b^2)^{1/2}]}{b^2}\},
\end{equation}
where $G$ is the gravitational constant, $M_L$ is the mass of
Laniakea, $d$ and $r_L=79\,$Mpc are the distances from the source and earth
to the center of Laniakea, $b$ is the impact parameter of the
particle paths relative to the center of Laniakea, and $s_n=+1$ or
$-1$ where the source is located along the direction or
anti-direction of Laniakea. Considering the center of the Great
Attractor \citep{Lynden-Bell1988ApJ...326...19L}, RA$_L=10^h32^m$ and
Dec$_L=-46^\circ00^\prime$ (the red point in top panel of figure
\ref{fig:distribution}) as the direction of Laniakea center, it has
\begin{equation}\label{eq:b}
b=r_L\sqrt{1-(\sin{\delta_s}\sin{\delta_L}+\cos{\delta_s}\cos{\delta_L}\cos{(\beta_s-\beta_L)})^2},
\end{equation}
where RA$_s=\delta_s$, Dec$_s=\beta_s$ and $\beta_L=10^h32^m$,
$\delta_L=-46^\circ00^\prime$ are the positions of the source and
the center of Laniakea respectively. With the data in our sample,
the value of $f(\mathbf{r})$ is about $10^{12}\,\rm s$.

Next we constrain the parameters $\Delta\gamma$, $\langle\Delta
t_{\rm other}\rangle$ by fitting the GRB time delay data in our
sample. Since the function $f(\mathbf{r})$ only depends on the angle
$\theta$ between the anti-direction of the center of Laniakea and
the direction of the source, we bin the time delay data though the
values of
$\mu=\cos(\theta)=\sin{\delta_s}\sin{\delta_L}+\cos{\delta_s}\cos{\delta_L}\cos{(\beta_s-\beta_L)}$.
To make sure that there are more bins and more data in each bin, we
choose the number of bins is $N = 20$.  In each bin $i$, we
calculate the mean time delay $\langle\Delta t_{\rm
obs}\rangle_{{\rm bin},i}$ and its uncertainty $\sigma_{{\rm
bin},i}={\rm std}(\Delta t_{\rm obs})/\sqrt{N_{{\rm bin},i}}$, where
${\rm std}(\Delta t_{\rm obs})$ and $N_{{\rm bin},i}$ are the
standard derivation of observed time delay and the number of data in
that bin. Besides, for each bin, we use the middle value of $\mu$
and average redshift as the values of that bin. These averaged data
are listed in table \ref{tab:binData}. Then we can fit the equation
(\ref{eq:fitFun2}) with the binned data to obtain the optimal
parameters.

\begin{figure}
    \centering
    \includegraphics[width=0.46\textwidth]{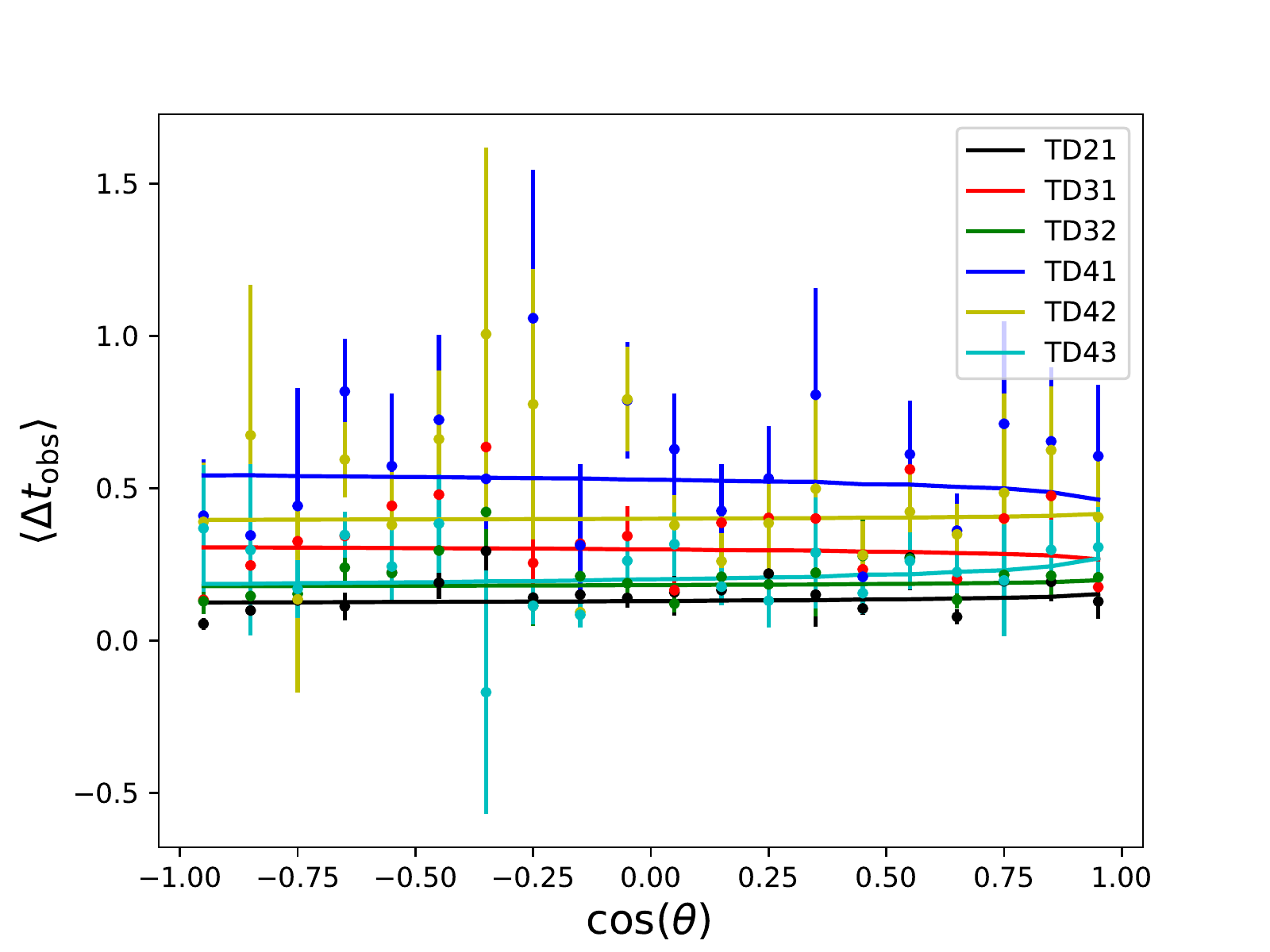}\\
    \includegraphics[width=0.46\textwidth]{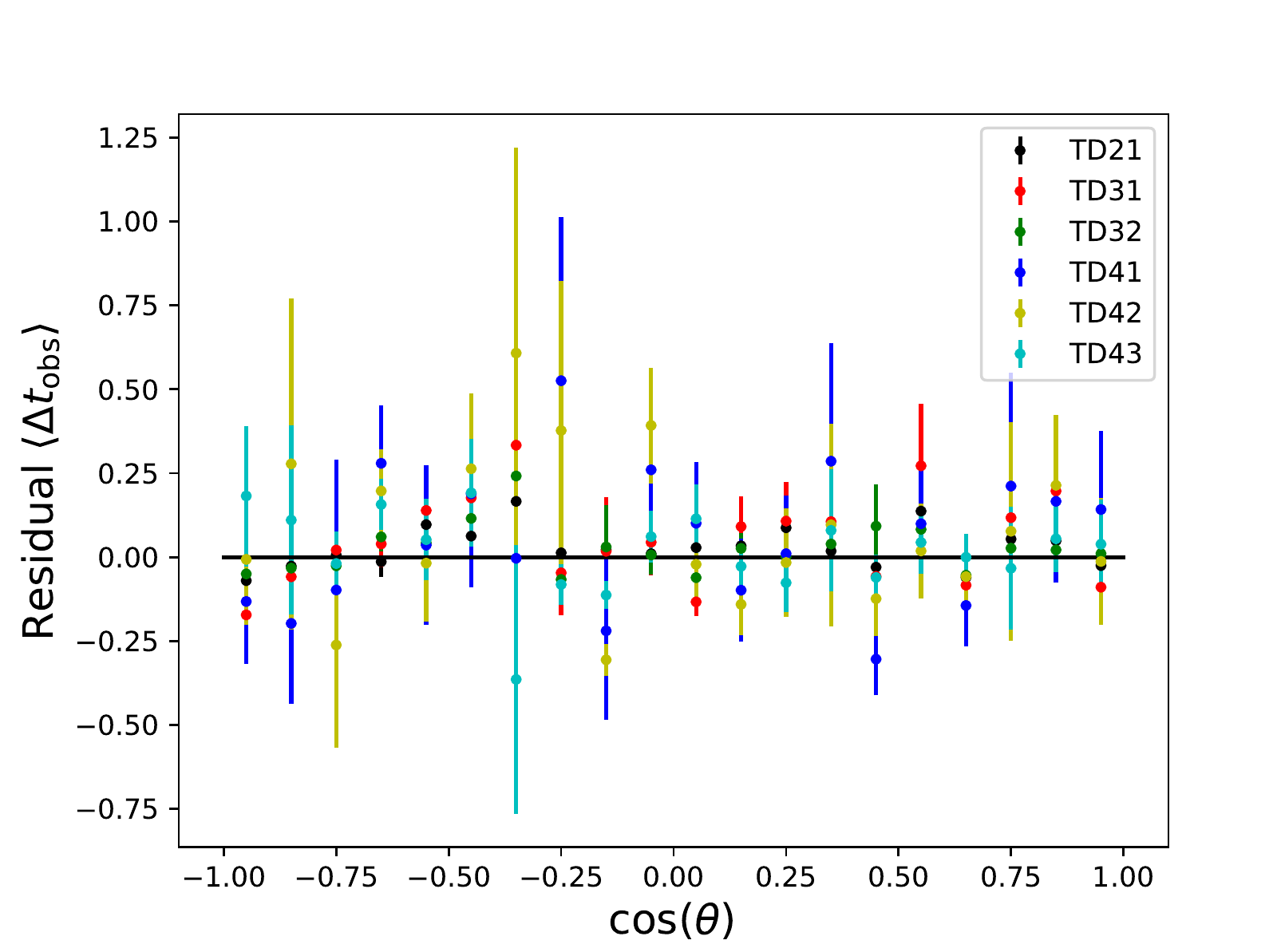}\\
    \caption{The top panel shows the optimal fit lines for the data and the bottom panel shows the residual
    of the observed time delay. Different colors represent different subsamples.}\label{fig:Fit}
\end{figure}

To describe the potential extra variance, we use the likelihood function proposed by \cite{DAgostini2005physics..11182D}
\begin{equation}\label{eq:likelihood}
\ln{\mathcal{L}} = -\frac{1}{2}\sum{\left[ \ln{(\sigma_{{\rm bin},i}^2+\sigma_{\rm extra}^2)}
+\frac{(\langle\Delta t_{\rm obs}\rangle_{{\rm bin},i}-\langle\Delta t_{\rm th}\rangle_{{\rm bin},i})^2}{\sigma_{{\rm bin},i}^2
+\sigma_{\rm extra}^2}\right] },
\end{equation}
where $\sigma_{\rm extra}$ is the extra potential uncertainty of the
binned data caused by the approximation. In this work, we use
$emcee$ \citep{Foreman-Mackey2013PASP..125..306F}, a Python Markov
chain Monte Carlo (MCMC) module to get the optimal values and
uncertainties of the parameters. Figure \ref{fig:Ch21} shows the
marginalized surfaces of the likelihood functions of the parameters
fitted with the TD21 (time delay between Ch1 and Ch2) data in our
sample. From this figure, we have
$\Delta\gamma_{21}=(0.02\pm0.03)\times10^{-12}$, $\langle\Delta
t_{\rm other,21}\rangle=0.09\pm0.09$ s and $\sigma_{\rm
extra,21}=0.04\pm0.02$ s with $1 \sigma$ uncertainty, which means
there is no evidence for violation of WEP. We have also used other 5
time delay data to constrain the WEP and the results are shown in
table \ref{tab:result}. From this table, we find that none of the 6
time delay data shows significant evidence of WEP violation. In
order to check whether the equation (\ref{eq:fitFun2}) can fit the
data well, we also plot the fit-lines of the data and the residual
values for different subsamples in Figure \ref{fig:Fit}, from which
we see the equation (\ref{eq:fitFun2}) can describe the binned data
well.

\section{Discussion and Summary}\label{sec:discussion}
Because of the importance of the WEP to the general relativity and
other metric gravity theory, many work have been done to test the
validity of WEP
\citep{Shapiro1964PhRvL..13..789S,Krauss1988PhRvL..60..176K,Longo1988PhRvL..60..173L,Gao2015ApJ...810..121G,
    Wei2015PhRvL.115z1101W,Nusser2016ApJ...821L...2N,Zhang2016arXiv160104558Z,Yang2016PhRvD..94j1501Y,Wang2016PhRvL.116o1101W,
    Wu2016PhRvD..94b4061W,Kahya2016PhLB..756..265K}. However, due to the
contamination of other effects, such as the $\Delta t_{\rm int}$,
$\Delta t_{\rm LIV}$ and $\Delta t_{\rm spec}$, in the observed time
delay $\Delta t_{\rm obs}$, they can only give the upper limit of
the violation of WEP in principle even if there is indeed some
violation of WEP. To solve this problem, we develop a new robust
method based on the global fitting of time delay measurement of
cosmic transients to constrain the violation of WEP. This is the
first method which can naturally subtract the effects of other terms
and give a confidence region of WEP violation.

\begin{figure}
    \centering
    \includegraphics[width=0.50\textwidth]{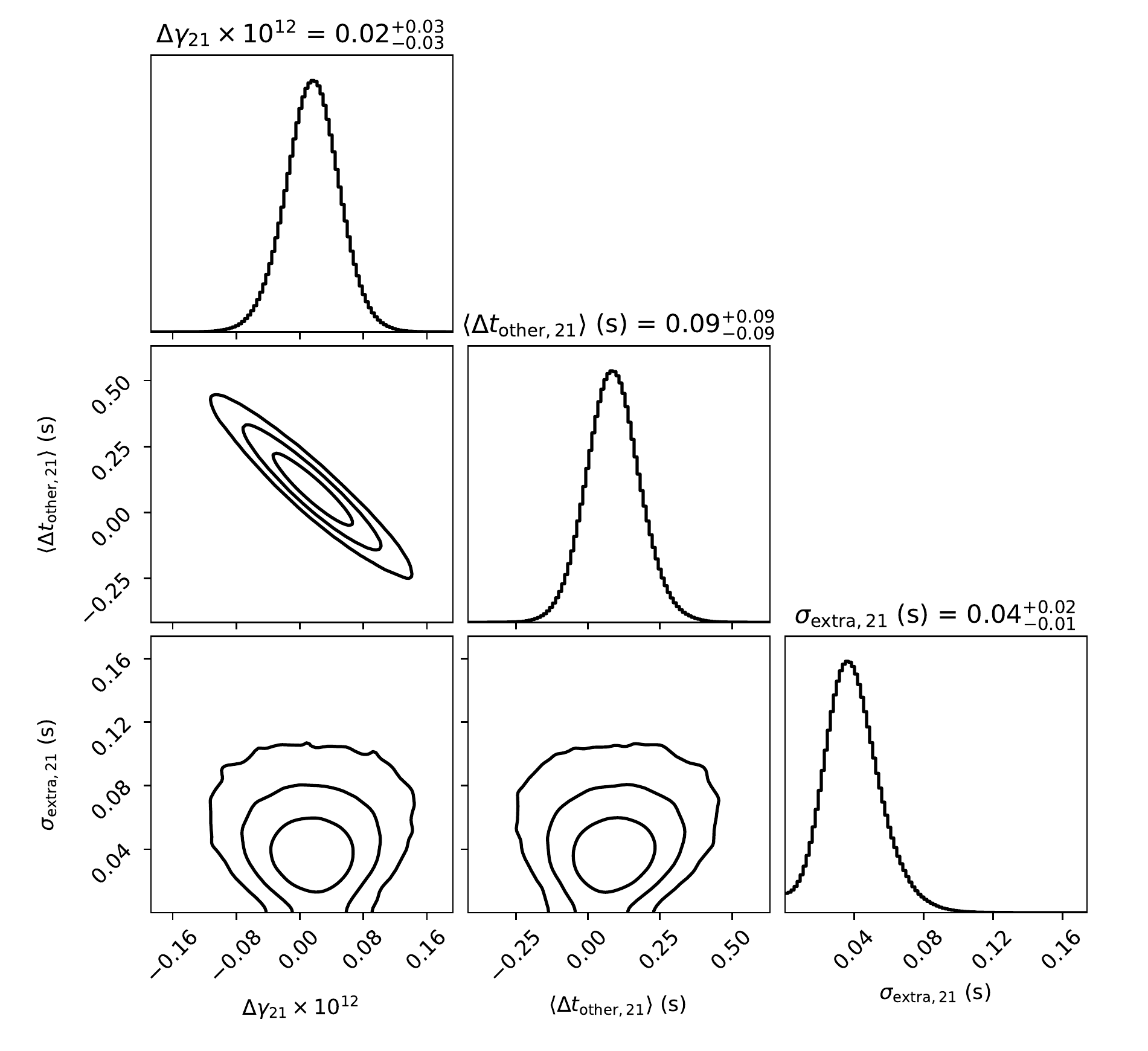}\\
    \caption{The surfaces of the marginalized likelihood functions of $\Delta\gamma$,
    $\left\langle \Delta t_{\rm other}\right\rangle $ and $\sigma_{\rm extra}$ derived from the TD21 data in our sample.}\label{fig:Ch21}
\end{figure}

We choose the time delay measurement data \citep{Hakkila2007ApJS..169...62H}
and the pseudo-redsihfts \citep{Yonetoku2004ApJ...609..935Y}
of 668 BATSE GRBs from previous literature. By using these data with our method,
we constrain the potential violation of WEP. The results are shown in figure
\ref{fig:Ch21} and table \ref{tab:result}. All time delay samples,
except TD43, show the violation of WEP is less than $1 \sigma$
confidence level. TD43 shows the largest violation of WEP at about
$2.5 \sigma$ confidence level. Therefore, there is no significant evidence for
WEP violation.

\begin{figure}
    \centering
    \includegraphics[width=0.5\textwidth]{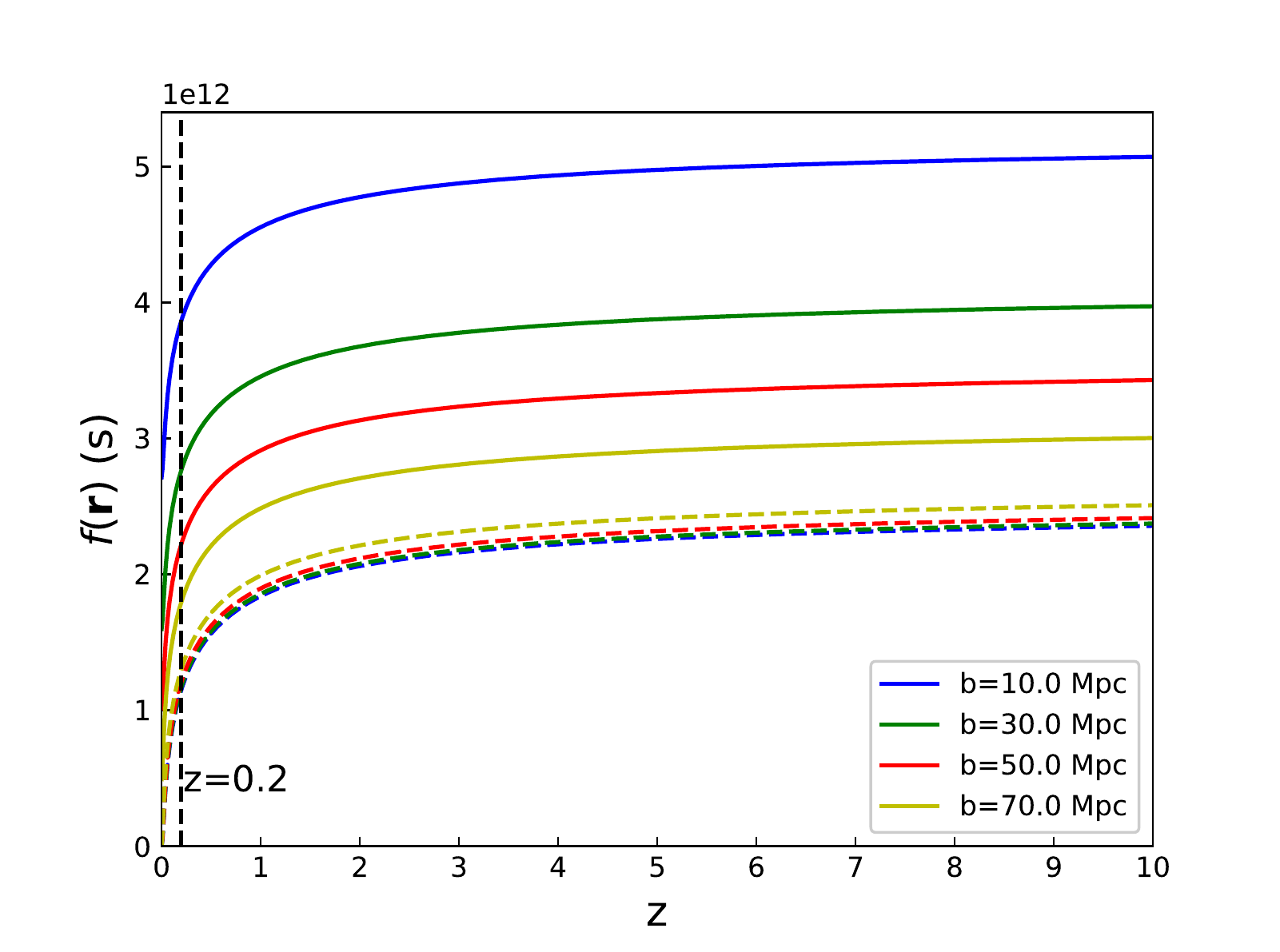}\\
    \caption{The value of $ f(\mathbf{r})$ as a function of redshift depends on different values of $b$
    and $s_n$ which represent the direction of the source. Different colors mean different values of $b$
    while the solid and dash lines mean $s_n=+1$ and $s_n=-1$ respectively.}\label{fig:fr}
\end{figure}

There are several uncertainties are not considered in the
analysis above. We will give a detailed discussion here which will
show that those uncertainties are not significant and can be totally
omitted in the data analysis process. Actually, the ability of our
method depends on the magnitude of $f(\mathbf{r})$ which is equation
(\ref{eq:fr}) and we can calculate the its uncertainty caused by the
uncertainty of the distance of GRBs.
\begin{equation}\label{eq:delta_fr}
\frac{\Delta f(\mathbf{r})}{f(\mathbf{r})} = \frac{ f^\prime(\mathbf{r})\Delta \mathbf{r}}{f(\mathbf{r})}
= \frac{\Delta d + d\Delta d/(d^2-b^2)^{1/2}}{d+(d^2-b^2)^{1/2}}/\ln\{\frac{[d+(d^2-b^2)^{1/2}][r_L+s_n(r_L^2-b^2)^{1/2}]}{b^2}\}
 \approx\frac{\Delta d}{d\ln{(d/b)}}\approx\frac{\Delta d}{4d},
\end{equation}
where we use the approximations $b\sim10\,$Mpc and
$d\sim10^3\,$Mpc.  Then let's see the effect of the uncertainty of
the redshift of GRBs. The comoving distance of a GRB at $z=0.2$ is
about $10^3\,$Mpc while that at $z=10$ is about $10^4\,$Mpc which is
only about 10 times further. Therefore, it can only affect
$f(\mathbf{r})$ by about 2 times and also about 2 times on the
confidence region of $\Delta\gamma$, which means there is no
significant effect even though we use very biased redshifts of GRBs.
We also drawn the values of $f(\mathbf{r})$ as a function of
redshift on figure \ref{fig:fr} based on different values of $b$ and
$s_n$ which depend only on the directions of GRBs. From figure
\ref{fig:fr}, we can also see that the uncertainty of redshift of
GRBs can not affect much on the value of $f(\mathbf{r})$ as long as
the GRBs are at cosmic distance. The other uncertainty is that the
energy channels we used in our work is not in the rest-frame so the
photons in a same energy channel will have much different energy
when they were emitted. However, from figure \ref{fig:fr} one can
find that the low redshift range, $z<0.2$, contributes most part of
the Shapiro time delay which is up to about 50\%-80\% depends on the
direction of the source. In this low redshift range, the photons in
same energy channel have similar energy. Therefore, the Shapiro time
delays for the photons in same observed frame energy channel are
caused at similar energy range.

The efficiency of our method depends on the sample size $N$ and the
extra variance $\sigma_{\rm int}$ of data. A larger $N$ and smaller
$\sigma_{\rm int}$ will give a tighter constraint on $\Delta\gamma$.
Since there are more than 2000 GRBs in BATSE sample and we here only
use 668 GRBs with pseudo-redshifts, a detailed analysis on the total
sample of BATSE GRBs sample may give a better constraint on it. Besides,
there will be more GRBs in Fermi GRB sample in the future.
Therefore, a systematical analysis on the time delay of Fermi GRBs
may also give a better constraint on this issue. The efficiency of our method also
depends on the position of observer in the local gravitational
potential and the directional distribution of the cosmic transients.
This can be represent by the relative variance of $f(\mathbf{r})$ of
the data sample $\frac{Var[f(\mathbf{r})]}{E[f(\mathbf{r})]}$, where
the $Var[f(\mathbf{r})]$ and $E[f(\mathbf{r})]$ are the variance and
expectation of $f(\mathbf{r})$ of the data sample. A sample with
larger value of $\frac{Var[f(\mathbf{r})]}{E[f(\mathbf{r})]}$ will
give a tighter constraint on $\Delta\gamma$.

Our method needs a large sample of time delay measurement.
Although GRBs are used in this work, other cosmic transients, such
as GW events and supernovae, are also attractive. Interestingly,
FRBs can be also a powerful tool to constrain WEP, if the $\Delta t_{\rm
    DM}$ could be subtracted.

\acknowledgments We thank the anonymous referee for detailed and
very constructive suggestions that have allowed us to improve our
manuscript. This work is supported by the National Basic Research
Program of China (973 Program, grant No. 2014CB845800), the National
Natural Science Foundation of China (grant Nos. 11422325 and 11373022),
and the Excellent Youth Foundation of Jiangsu
Province (BK20140016).

\bibliographystyle{aasjournal}
\bibliography{bibfile}

\begin{thebibliography}{}
\expandafter\ifx\csname natexlab\endcsname\relax\def\natexlab#1{#1}\fi
\providecommand{\url}[1]{\href{#1}{#1}}

\bibitem[{{Amelino-Camelia}(2013)}]{Amelino-Camelia2013LRR....16....5A}
{Amelino-Camelia}, G. 2013, Living Reviews in Relativity, 16, 5

\bibitem[{{D'Agostini}(2005)}]{DAgostini2005physics..11182D}
{D'Agostini}, G. 2005, ArXiv Physics e-prints, physics/0511182

\bibitem[{{Foreman-Mackey} {et~al.}(2013){Foreman-Mackey}, {Hogg}, {Lang}, \&
  {Goodman}}]{Foreman-Mackey2013PASP..125..306F}
{Foreman-Mackey}, D., {Hogg}, D.~W., {Lang}, D., \& {Goodman}, J. 2013, \pasp,
  125, 306

\bibitem[{{Gao} {et~al.}(2015){Gao}, {Wu}, \&
  {M{\'e}sz{\'a}ros}}]{Gao2015ApJ...810..121G}
{Gao}, H., {Wu}, X.-F., \& {M{\'e}sz{\'a}ros}, P. 2015, \apj, 810, 121

\bibitem[{{Goldstein} {et~al.}(2013){Goldstein}, {Preece}, {Mallozzi},
  {Briggs}, {Fishman}, {Kouveliotou}, {Paciesas}, \&
  {Burgess}}]{Goldstein2013ApJS..208...21G}
{Goldstein}, A., {Preece}, R.~D., {Mallozzi}, R.~S., {et~al.} 2013, \apjs, 208,
  21

\bibitem[{{Hakkila} {et~al.}(2007){Hakkila}, {Giblin}, {Young}, {Fuller},
  {Peters}, {Nolan}, {Sonnett}, {Haglin}, \&
  {Roiger}}]{Hakkila2007ApJS..169...62H}
{Hakkila}, J., {Giblin}, T.~W., {Young}, K.~C., {et~al.} 2007, \apjs, 169, 62

\bibitem[{{Kahya} \& {Desai}(2016)}]{Kahya2016PhLB..756..265K}
{Kahya}, E.~O., \& {Desai}, S. 2016, Physics Letters B, 756, 265

\bibitem[{{Krauss} \& {Tremaine}(1988)}]{Krauss1988PhRvL..60..176K}
{Krauss}, L.~M., \& {Tremaine}, S. 1988, Physical Review Letters, 60, 176

\bibitem[{{Kumar} \& {Zhang}(2015)}]{Kumar2015PhR...561....1K}
{Kumar}, P., \& {Zhang}, B. 2015, \physrep, 561, 1

\bibitem[{{Longo}(1988)}]{Longo1988PhRvL..60..173L}
{Longo}, M.~J. 1988, Physical Review Letters, 60, 173

\bibitem[{{Lynden-Bell} {et~al.}(1988){Lynden-Bell}, {Faber}, {Burstein},
  {Davies}, {Dressler}, {Terlevich}, \&
  {Wegner}}]{Lynden-Bell1988ApJ...326...19L}
{Lynden-Bell}, D., {Faber}, S.~M., {Burstein}, D., {et~al.} 1988, \apj, 326, 19

\bibitem[{{M{\'e}sz{\'a}ros}(2006)}]{Meszaros2006RPPh...69.2259M}
{M{\'e}sz{\'a}ros}, P. 2006, Reports on Progress in Physics, 69, 2259

\bibitem[{{Nusser}(2016)}]{Nusser2016ApJ...821L...2N}
{Nusser}, A. 2016, \apjl, 821, L2

\bibitem[{{Shapiro}(1964)}]{Shapiro1964PhRvL..13..789S}
{Shapiro}, I.~I. 1964, Physical Review Letters, 13, 789

\bibitem[{{Tully} {et~al.}(2014){Tully}, {Courtois}, {Hoffman}, \&
  {Pomar{\`e}de}}]{Tully2014Natur.513...71T}
{Tully}, R.~B., {Courtois}, H., {Hoffman}, Y., \& {Pomar{\`e}de}, D. 2014,
  \nat, 513, 71

\bibitem[{{Wang} {et~al.}(2015){Wang}, {Dai}, \&
  {Liang}}]{Wang2015NewAR..67....1W}
{Wang}, F.~Y., {Dai}, Z.~G., \& {Liang}, E.~W. 2015, \nar, 67, 1

\bibitem[{{Wang} {et~al.}(2016){Wang}, {Liu}, \&
  {Wang}}]{Wang2016PhRvL.116o1101W}
{Wang}, Z.-Y., {Liu}, R.-Y., \& {Wang}, X.-Y. 2016, Physical Review Letters,
  116, 151101

\bibitem[{{Wei} {et~al.}(2015){Wei}, {Gao}, {Wu}, \&
  {M{\'e}sz{\'a}ros}}]{Wei2015PhRvL.115z1101W}
{Wei}, J.-J., {Gao}, H., {Wu}, X.-F., \& {M{\'e}sz{\'a}ros}, P. 2015, Physical
  Review Letters, 115, 261101

\bibitem[{{Wei} {et~al.}(2017){Wei}, {Zhang}, {Shao}, {Wu}, \&
  {M{\'e}sz{\'a}ros}}]{Wei2017ApJ...834L..13W}
{Wei}, J.-J., {Zhang}, B.-B., {Shao}, L., {Wu}, X.-F., \& {M{\'e}sz{\'a}ros},
  P. 2017, \apjl, 834, L13

\bibitem[{{Will}(2014)}]{Will2014LRR....17....4W}
{Will}, C.~M. 2014, Living Reviews in Relativity, 17, 4

\bibitem[{{Wu} {et~al.}(2016){Wu}, {Gao}, {Wei}, {M{\'e}sz{\'a}ros}, {Zhang},
  {Dai}, {Zhang}, \& {Zhu}}]{Wu2016PhRvD..94b4061W}
{Wu}, X.-F., {Gao}, H., {Wei}, J.-J., {et~al.} 2016, \prd, 94, 024061

\bibitem[{{Yang} {et~al.}(2017){Yang}, {Zou}, {Zhang}, {Liao}, \&
  {Lei}}]{2017MNRAS.469L..36Y}
{Yang}, C., {Zou}, Y.-C., {Zhang}, Y.-Y., {Liao}, B., \& {Lei}, W.-H. 2017,
  \mnras, 469, L36

\bibitem[{{Yang} \& {Zhang}(2016)}]{Yang2016PhRvD..94j1501Y}
{Yang}, Y.-P., \& {Zhang}, B. 2016, \prd, 94, 101501

\bibitem[{{Yonetoku} {et~al.}(2004){Yonetoku}, {Murakami}, {Nakamura},
  {Yamazaki}, {Inoue}, \& {Ioka}}]{Yonetoku2004ApJ...609..935Y}
{Yonetoku}, D., {Murakami}, T., {Nakamura}, T., {et~al.} 2004, \apj, 609, 935

\bibitem[{{Yu} \& {Wang}(2018)}]{Yu2018arXiv180101257Y}
{Yu}, H., \& {Wang}, F.~Y. 2018, ArXiv e-prints, arXiv:1801.01257

\bibitem[{{Zhang}(2016)}]{Zhang2016arXiv160104558Z}
{Zhang}, S.-N. 2016, ArXiv e-prints, arXiv:1601.04558

\end{thebibliography}

\begin{center}
\setlength{\tabcolsep}{2pt}
\begin{longtable}{c|c|c|c|c|c|c|c|c|c}
         \caption{The time delay data used in this work.}
            \label{tab:data}  \\
            \hline
GRB & $z$ & RA & DEC & TD21 (s) & TD31 (s) &  TD32 (s) & TD41 (s) & TD42 (s) & TD43 (s) \\ \hline
\endfirsthead

\hline
GRB & $z$ & RA & DEC & TD21 (s) & TD31 (s) &  TD32 (s) & TD41 (s) & TD42 (s) & TD43 (s) \\ \hline
\endhead

\hline
\endfoot

105 &   1.26    $^{+    0.11    }_{-    0.10    }$ &    270.68  &   24.76   &   0.030   $\pm$   0.013   &   0.026   $\pm$   0.030   &   -0.005  $\pm$   0.011   &   -0.064  $\pm$   0.001   &   -0.091  $\pm$   0.034   &   -0.064  $\pm$   0.001   \\
109 &   1.98    $^{+    0.20    }_{-    0.17    }$ &    91.29   &   -22.77  &   0.041   $\pm$   0.014   &   0.022   $\pm$   0.031   &   0.021   $\pm$   0.010   &   0.087   $\pm$   0.142   &   -0.013  $\pm$   0.036   &   0.008   $\pm$   0.009   \\
111 &   11.23   $^{+    33.87   }_{-    6.51    }$ &    75.82   &   -19.51  &   2.692   $\pm$   0.134   &   5.941   $\pm$   0.086   &   3.371   $\pm$   0.152   &   None    &   None    &   None    \\
130 &   5.17    $^{+    0.68    }_{-    0.57    }$ &    135.8   &   2.73    &   0.019   $\pm$   0.006   &   0.035   $\pm$   0.008   &   0.021   $\pm$   0.002   &   -0.042  $\pm$   0.013   &   -0.037  $\pm$   0.018   &   -0.041  $\pm$   0.008   \\
133 &   4.53    $^{+    9.47    }_{-    2.89    }$ &    126.54  &   -0.23   &   0.936   $\pm$   0.178   &   1.386   $\pm$   0.260   &   0.410   $\pm$   0.082   &   2.651   $\pm$   0.390   &   0.638   $\pm$   1.637   &   None    \\
143 &   1.36    $^{+    0.04    }_{-    0.03    }$ &    87.45   &   38.74   &   0.019   $\pm$   0.004   &   0.042   $\pm$   0.001   &   0.030   $\pm$   0.002   &   0.090   $\pm$   0.014   &   0.071   $\pm$   0.005   &   0.036   $\pm$   0.003   \\
160 &   4.39    $^{+    1.10    }_{-    0.78    }$ &    299.55  &   -44.88  &   0.012   $\pm$   0.003   &   0.057   $\pm$   0.013   &   0.002   $\pm$   0.004   &   -0.331  $\pm$   0.055   &   -0.213  $\pm$   0.064   &   0.000   $\pm$   0.001   \\
179 &   8.46    $^{+    8.12    }_{-    3.41    }$ &    266.48  &   58.3    &   0.021   $\pm$   0.015   &   0.049   $\pm$   0.032   &   0.035   $\pm$   0.007   &   0.062   $\pm$   0.069   &   0.039   $\pm$   0.035   &   0.070   $\pm$   0.047   \\
\end{longtable}
\end{center}

\begin{table}[!hbp]
    \begin{center}
        \begin{tabular}{c|c|c|c|c|c|c}
            \hline

            Sample  &   Statistical property &  Total   &   NE  &   NW  &   SE  &   SW  \\\hline
            &   Size    &   668 &   179 &   164 &   167 &   158 \\
            TD21    &   Mean    &   0.16    &   0.15    &   0.15    &   0.18    &   0.14    \\
            &   STD &   0.37    &   0.35    &   0.38    &   0.39    &   0.34    \\
            &   KS-test p-value &   1.00    &   0.89    &   0.98    &   0.93    &   0.72    \\\hline
            &   Size    &   664 &   178 &   163 &   167 &   156 \\
            TD31    &   Mean    &   0.34    &   0.34    &   0.33    &   0.41    &   0.28    \\
            &   STD &   0.67    &   0.66    &   0.64    &   0.8 &   0.52    \\
\hline

        \end{tabular}
        \caption{The statistical properties of GRBs in four different areas of the sky and the whole sample. The KS-test p-values between the subsamples and the whole samples show that they are sampled from same distributions.}\label{tab:statistic}
    \end{center}
\end{table}

\begin{table}[!hbp]
    \begin{center}
        \setlength{\tabcolsep}{2pt}
        \begin{tabular}{c|c|c|c|c|c|c|c|c|c|c|c|c|c}
            \hline
            \multirow{2}*{Bin's ID}  & \multirow{2}*{$\mu$} & \multicolumn{2}{|c|}{TD21} & \multicolumn{2}{|c|}{TD31} & \multicolumn{2}{|c|}{TD32} & \multicolumn{2}{|c|}{TD41} & \multicolumn{2}{|c|}{TD42} & \multicolumn{2}{|c}{TD43}  \\
            \cline{3-14}
            ~ & ~ & $\left\langle z\right\rangle $ & $\langle\Delta t_{\rm obs}\rangle_{\rm bin}$ (s) & $\left\langle z\right\rangle $ & $\langle\Delta t_{\rm obs}\rangle_{\rm bin}$ (s) & $\left\langle z\right\rangle $ & $\langle\Delta t_{\rm obs}\rangle_{\rm bin}$ (s) & $\left\langle z\right\rangle $ & $\langle\Delta t_{\rm obs}\rangle_{\rm bin}$ (s) & $\left\langle z\right\rangle $ & $\langle\Delta t_{\rm obs}\rangle_{\rm bin}$ (s) & $\left\langle z\right\rangle $ & $\langle\Delta t_{\rm obs}\rangle_{\rm bin}$ (s)  \\ \hline

            1 & -0.95 & 4.29 & 0.06$\pm$0.02 & 4.29 & 0.13$\pm$0.03 & 4.20 & 0.13$\pm$0.04 & 4.19 & 0.41$\pm$0.19 & 4.22 & 0.39$\pm$0.20 & 4.03 & 0.37$\pm$0.21 \\
            2 & -0.85 & 4.11 & 0.10$\pm$0.03 & 4.11 & 0.25$\pm$0.05 & 4.11 & 0.15$\pm$0.04 & 3.39 & 0.34$\pm$0.24 & 3.68 & 0.67$\pm$0.49 & 3.81 & 0.30$\pm$0.28 \\
            3 & -0.75 & 3.85 & 0.13$\pm$0.06 & 3.85 & 0.33$\pm$0.11 & 3.85 & 0.15$\pm$0.04 & 4.12 & 0.44$\pm$0.39 & 4.39 & 0.14$\pm$0.31 & 4.10 & 0.17$\pm$0.09 \\
            4 & -0.65 & 3.92 & 0.11$\pm$0.04 & 3.92 & 0.34$\pm$0.07 & 3.92 & 0.24$\pm$0.04 & 4.10 & 0.82$\pm$0.17 & 4.10 & 0.59$\pm$0.12 & 3.94 & 0.35$\pm$0.07 \\
            5 & -0.55 & 4.18 & 0.22$\pm$0.07 & 4.18 & 0.44$\pm$0.13 & 3.99 & 0.22$\pm$0.07 & 4.14 & 0.57$\pm$0.24 & 3.94 & 0.38$\pm$0.17 & 3.94 & 0.24$\pm$0.12 \\
            6 & -0.45 & 4.07 & 0.19$\pm$0.05 & 4.07 & 0.48$\pm$0.12 & 4.07 & 0.29$\pm$0.06 & 3.83 & 0.72$\pm$0.28 & 4.03 & 0.66$\pm$0.22 & 3.69 & 0.38$\pm$0.16 \\
\hline

        \end{tabular}
        \caption{The binned data which are used to fit.}\label{tab:binData}
    \end{center}
\end{table}

\begin{table}[!hbp]
    \begin{center}
        \begin{tabular}{c|c|c|c|c}
            \hline
            Data & Sample size & $\Delta\gamma (10^{-12})$ & $\left\langle \Delta t_{\rm other}\right\rangle  (s)$ & $\sigma_{\rm extra} (s)$ \\ \hline
            TD21 & 668 & $0.02\pm0.03$ & $0.09\pm0.09$ & $0.04\pm0.02$ \\ \hline
            TD31 & 664 & $-0.02\pm0.06$ & $0.35\pm0.18$ & $0.03\pm0.02$ \\ \hline
            TD32 & 662 & $0.01\pm0.04$ & $0.16\pm0.10$ & $0.03\pm0.02$ \\ \hline
            TD41 & 449 & $-0.05\pm0.13$ & $0.64\pm0.35$ & $0.11\pm0.07$  \\ \hline
            TD42 & 423 & $0.01\pm0.12$ & $0.36\pm0.35$ & $0.16\pm0.05$  \\ \hline
            TD43 & 387 & $0.05\pm0.07$ & $0.08\pm0.18$ & $0.06\pm0.03$  \\ \hline
        \end{tabular}
        \caption{Sample size and the optimal values with 1 $\sigma$ uncertainties of the parameters $\Delta\gamma$, $\Delta t_{\rm other}$ and $\sigma_{\rm extra}$ with
            time delay measurement data of GRBs in our sample. TD$ij$ means the number of the time delay measurements between ${\rm Ch}i$ and ${\rm Ch}j$.}\label{tab:result}
    \end{center}
\end{table}

\end{document}